\documentstyle[aps,epsf,floats,twocolumn]{revtex}
\renewenvironment{abstract}[1]{\vspace*{-6ex}
  \hspace*{0.6in}\begin{minipage}[t][#1]{5.5in}\small}
  {\end{minipage}}

\newcommand{\twocolbreakpar}[1]{\newline \\[#1] \par \noindent}
\begin{document}
\draft
\title{Molecular Dynamics Simulation of Binary Hard-Sphere
	Crystal/Melt Interfaces}
\author{Ruslan L. Davidchack and Brian. B. Laird\cite{Brian}}
\address{Department of Chemistry and Kansas Institute for 
	 Theoretical and Computational Science,\\
         University of Kansas, Lawrence, Kansas 66045, USA}
\date{\today}
\maketitle

\begin{abstract}{1.1in}
We examine, using molecular dynamics simulation, the structure
and thermodynamics of the (100) and (111) disordered face-centered 
cubic (FCC) crystal/melt interfaces for a binary hard-sphere system. 
This study is an extension of our previous work, 
[Phys. Rev. E {\bf 54}, R5905 (1996)], in which preliminary data for 
the (100) interface were reported.  Density and diffusion profiles on 
both fine- and course-grained scales are calculated and analyzed 
leading to the conclusion that equilibrium interfacial segregation is 
minimal in this system. 
\end{abstract}

\section{Introduction}
In order to fully understand such important phenomena as 
near-equilibrium crystal growth, homogeneous nucleation and 
interfacial solute segregation, a detailed microscopic description 
of the crystal/melt interface is 
necessary\cite{Woodruff73,Tiller91,Howe97}. 
Since direct experimental data is scarce, due to the extreme 
difficulty of constructing and interpreting experiments on such 
systems, computer simulation has become an important tool, not only 
in its usual role of aiding in the development and evaluation of 
interface theories\cite{Singh91,Lowen94,Laird92b}, but also in
determining the basic microscopic phenomenology of crystal/melt 
interfaces.  Previous simulation studies of such interfaces have
focused entirely on single-component systems (for a review of recent 
simulation studies, see Ref. \cite{Laird92b}); however, since 
most technologically important materials are not pure substances but 
mixtures (such as alloys), it is important that such studies be 
extended to multi-component interfacial systems.  In addition to the
usual issues of interfacial width, interfacial free energy, and
transport within the interface, multi-component systems allow for
the study of interfacial segregation.

As in single-component systems, the crystal/melt interface of a 
multi-component system is characterized by measuring the change in the
various structural, thermal and dynamical properties of interest as
one traverses the interface from one bulk phase into the other.
For planar interfaces (the type studied here), the $z$ axis is usually
taken as the direction perpendicular to the interfacial plane and
quantities are averaged over $x$ and $y$ and presented as functions
of $z$.  Examples include the density profiles of the various
components (labeled by $i$), 
$\rho_{i}(z) = \langle\rho_{i}({\bf r})\rangle_{xy}$
and the diffusion constant profiles, $D_{i}(z)$.  The thermodynamic 
quantity of most interest is the solid-liquid interfacial free energy,
$\gamma_{\mathrm{sl}}$, which is defined as the work required to form 
one unit area of surface -- this quantity is extremely difficult to 
calculate via simulation -- even in the single-component case.  
The only reliable calculation is the calculation of 
$\gamma_{\mathrm{sl}}$ for a single-component Lennard-Jones system 
by Broughton and Gilmer\cite{Broughton86c}.
\twocolbreakpar{0.77in}

Recently\cite{Davidchack96}, we have reported initial simulation 
results on the crystal/melt interface of a two-component hard-sphere 
mixture.  Specifically, the interface between the (100) face of a 
disordered face-centered cubic (FCC) crystal and the coexisting 
fluid was studied.  In this work we extend this calculation to 
the (111) face and revisit the (100) interface calculations
using more detailed analysis.  These simulations represent the first
such simulations on the crystal/melt interface of a multi-component
system. 

The reasons for choosing the hard-sphere system for this initial
study are two-fold: First, it is now well established that the 
structure and freezing behavior of dense, simple fluids, is 
determined, for the most part, by packing considerations determined
by the repulsive part of the interaction potential.  The effect of
the attractive forces can generally be accounted for by treating
it as a perturbation to the repulsive part of the potential, which
is often approximated by a hard-sphere with some effective 
diameter\cite{Hansen86}.  Thus, the hard-sphere system is a useful 
reference from which to begin studies of more realistic systems.
Studying the hard-sphere system also allows one to directly probe 
the role of packing (which is purely entropic) in determining the 
interfacial phenomenology.  Second, the relative simplicity of the 
hard-sphere system lends itself well to theoretical study -- the 
vast majority of density-functional calculations on crystal/melt 
interfaces involve hard-sphere systems.  The disordered-FCC/fluid 
interfaces were chosen for initial study, because, in order to begin
an interface simulation, the phase coexistence conditions must
be very accurately known, and the disordered-FCC/fluid region of the
binary hard-sphere phase diagram (which occurs in the region of the
phase diagram where the difference between the diameters of the 
different components is not too great) has been well 
characterized\cite{Kranendonk89}.

In the next section we describe the binary hard-sphere system and
its phase behavior.  The results of the disordered FCC/fluid
interfacial simulations reported for the (100) and (111) 
interfaces are discussed in Section 3.  In Section 4, we conclude.

\section{The Binary Hard-Sphere System}
We consider a two-component system consisting of hard spheres of 
differing sizes.  The interaction potential for such a system can 
be written
\begin{equation}
\phi_{ij}(r) = \left \{ \begin{array}{ll}
\infty \quad & r < \sigma_{ij} \\
0  & r \ge \sigma_{ij}
\end{array}
\right . \;,
\end{equation}
where $r$ is the distance between the centers of two spheres 
and $i$ and $j\in \{1,2\}$ index the two types of spheres, which
are distinguished by their different diameters, $\sigma_1$ and 
$\sigma_2$, while their masses are assumed to be identical.  We also 
assume that the spheres are additive; i.e., 
$\sigma_{11} = \sigma_{1}$, $\sigma_{22} = \sigma_{2}$, 
and $\sigma_{12} = \sigma_{21} = (\sigma_{1} + \sigma_2)/2$.
The following definitions and conventions are adopted for this study.
First, it is assumed, without loss of generality, that
$\sigma_2 \geq \sigma_1$, and the hard-sphere {\em diameter ratio}
$\alpha$ is defined as
\begin{equation}
\alpha = \sigma_1/\sigma_2,
\quad (0 \leq \alpha \leq 1)\:.
						\label{eq:alpha}  
\end{equation} 
Second, if there are $N_1$ hard spheres with diameter $\sigma_1$
and $N_2$ with diameter $\sigma_2$ in the volume $V$, then
\begin{equation}
\rho = \frac{N_1+N_2}{V} = \rho_1 + \rho_2
						\label{eq:rho}  
\end{equation} 
is the {\em total} number density, and $\rho_i$'s represent the
respective number densities for individual species.  Third, the
concentrations (mole fractions) of individual components are given by
\begin{equation}
x_i = \rho_i/\rho, \quad i = 1,2\:.
						\label{eq:xi}  
\end{equation} 
Since $x_1 + x_2 = 1$, a single variable $x$ is usually used, such 
that $x_2 = x$ and $x_1 = 1-x$.

Also, the total packing fraction for the mixture $\eta$ is 
\begin{equation}
\eta = \eta_1 + \eta_2\:,
						\label{eq:eta}  
\end{equation} 
where $\eta_i = \frac{\pi}{6}\sigma_i^3\,\rho_i$, $i = 1,2$. 
The diameter ratio $\alpha$ together with any pair of independent 
parameters from those defined above can be used to completely 
specify the fluid state of a binary hard-sphere system.
The unit of length for the binary system is
taken to be equal to the diameter of a larger sphere $\sigma_2$.

Depending on the value of the diameter ratio $\alpha$, different 
crystal structures may have the lowest free energy 
\cite{Hume-Rothery69,Denton90}.  In particular, for $\alpha$ above 
about 0.85, the  {\em substitutionally disordered} FCC crystal, in 
which spheres of different diameters are distributed randomly over 
the sites of an FCC lattice, is the stable structure at freezing.  
For $\alpha$ below this value, a variety of structures may exist 
including ordered solid such as NaCl and CsCl 
structures\cite{Denton90}.  For an interface simulation study, it
is necessary that the phase coexistence conditions be accurately
known.  Therefore, we have chosen for this study the 
disordered-FCC/melt system, since the coexistence conditions for the 
equilibrium disordered crystal/melt interface can be found in the
study of Kranendonk and Frenkel\cite{Kranendonk89}, who have 
calculated the crystal/melt phase diagrams for several values 
of the diameter ratio in the range $0.85 < \alpha < 1$.  For
the present study, the diameter ratio $\alpha = 0.9$ is selected.
\begin{figure}[t]
\epsfxsize=9.4cm \hspace*{-8mm}\epsfbox{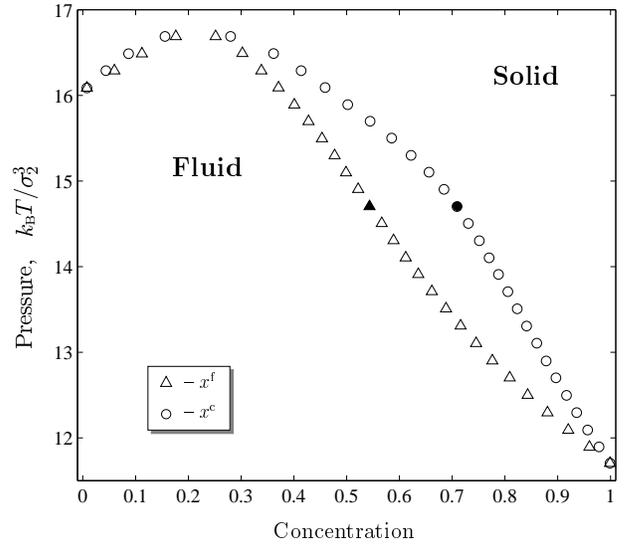} 
\vspace*{2ex}
\caption[Phase diagram of a binary hard-sphere system with 
$\alpha = 0.9\:.$]
{Phase diagram of a binary hard-sphere system with $\alpha = 0.9\:.$  
The data are taken from simulation by Kranendonk and 
Frenkel \protect\cite{Kranendonk89}.  The triangles correspond to 
the fluid phase, and circles to the crystal phase.  The solid 
triangle and circle represent the coexistence state selected for the 
interface simulation in the present work.}
\label{fig:kf}
\end{figure}   

For a binary system with a given diameter ratio $\alpha$, the
coexistence state is specified by two number densities
$\rho^{\mathrm c}$ and $\rho^{\mathrm f}$,
as well as two concentrations $x^{\mathrm c}$ and $x^{\mathrm f}$
(or, alternatively, by four number densities of individual species,
$\rho_i^{\mathrm c}$ and $\rho_i^{\mathrm f}$, where $i = 1,2$).
The pressure-concentration phase diagram determined in the
simulation by Kranendonk and
Frenkel\cite{Kranendonk89} for $\alpha = 0.9$ is shown 
in Fig.~\ref{fig:kf}.  At a
given pressure, the fluid and crystal phases have different 
concentrations at coexistence, represented in the plot by 
triangles and circles, respectively.  To maximize the 
deviation from single-component behavior, a point on the 
$\alpha = 0.9$ phase diagram was chosen for the interface simulation,
where the concentration difference between crystal and melt phases
is the largest.  This point occurs at a pressure of
\begin{equation}
P = 14.7\, k_{\mathrm B}  T/\sigma_2^3		\label{eq:pnum}
\end{equation}
and concentrations of 
\begin{equation}
x^{\mathrm c} = 0.71 \quad \mbox{and} 
\quad x^{\mathrm f} = 0.54 		\label{eq:xnum}
\end{equation}
for the crystal and fluid phase, respectively, 
and is shown in Fig. \ref{fig:kf} by the solid triangle and circle.
We have found the total packing fractions (number densities) in 
each two phases at
\begin{eqnarray}
  \eta^{\mathrm c}&=&0.552\qquad(\rho^{\mathrm c}\sigma_2^3 = 1.144)
  \:,\nonumber \\[-.15cm]
  \eta^{\mathrm f}&=&0.502\qquad(\rho^{\mathrm f}\sigma_2^3 = 1.096)
 				        \label{eq:binpar}
\end{eqnarray}  
and have used these values for the initialization of the binary 
interface systems.  We have run simulations for several trial
systems and have found that the systems are stable and the 
bulk crystal remains on average stress-free.  Therefore, the above
parameters have been used for all the simulation runs from which
average properties of the interface have been computed.

\section{Interface Construction and Simulation Results}
To create the initial bulk systems that are placed together to
form the interface, the two sphere types are distributed randomly 
according to the crystal and fluid coexistence concentrations.
The concentration in each crystal layer is maintained fixed by
randomly distributing the spheres of different types on a
layer by layer basis, thereby removing layer-to-layer concentration 
fluctuations due to finite system size.
This constraint is not expected to affect the results in any 
significant way besides removing the fluctuations
that cannot be averaged over during the simulation run due to 
practically no diffusion in the bulk crystal. The random distribution
of particle types is justified by the conclusion of Kranendonk
and Frenkel that above about $x = 0.6$ there is little or no local
substitutional ordering\cite{Kranendonk91b}.

Since the hard-sphere system evolves on a collision-by-collision 
basis, the natural unit of time for hard spheres is the mean collision
time $\tau_{\mathrm c}$, i.e. the average time between collisions
suffered by a given particle.  On the other hand, the duration of a 
simulation run is most conveniently measured in terms of the total 
number of collisions.  In the present study, in order to have better 
correspondence between the simulation time and the system evolution 
time, we measure simulation time in units 
of the number of {\em collisions per particle} (cpp),
defined as twice the ratio of the total number of 
collisions to the number of particles in the system, so that 
$ 1\:{\mathrm cpp} \:\approx\: \tau_{\mathrm c}\:$.
                                                
We have prepared 10 systems for each of (100) and (111) crystal
orientations and have run the simulations for 20\,000\,cpp with
the interfacial diagnostics being recorded every 200\,cpp.
The (100) systems contain 11\,616 spheres 
and have dimensions $L_x = L_y = 16.70\,\sigma_2$, and 
$L_z = 37.09\,\sigma_2$, while the size of the (111) systems 
is 11\,340 spheres and $L_x = 16.09\,\sigma_2$, 
$L_y = 16.72\,\sigma_2$, and $L_z = 37.60\,\sigma_2$.
(We have also done simulations on these systems with smaller
numbers of spheres -- $\sim$ 3000 and 6000 -- and different 
cross-sectional shapes.  The results of these simulations are
within statistical error, quantitatively identical to those
presented here.  The larger samples have been chosen as they give
better statistics, have much shorter equilibration times than the
smaller samples, and exhibit smaller interfacial fluctuations.) 

The concentration fluctuations in the fluid and the interfacial 
regions have been found to be much larger and more persistent than 
the density fluctuations.  Also, due to a finite system size,
even though the total momentum with respect to the simulation cell is
set to zero, drift of the interface positions is 
observed.  In order to avoid broadening of the
interfacial profiles caused by the drift, we have
selected for the final averages 12 segments of 2000\,cpp in duration 
from each crystal orientation, such that the drift does not exceed
half the distance between crystal layers (for more details on the
methods of  interfacial construction and equilibration used here, as
well as more detailed definitions of measured (computed) profiles, see
our recent work on single-component systems~\cite{Davidchack98}).    

\begin{figure}[t]
\epsfxsize=8.3cm  \hspace*{-2mm}\epsfbox{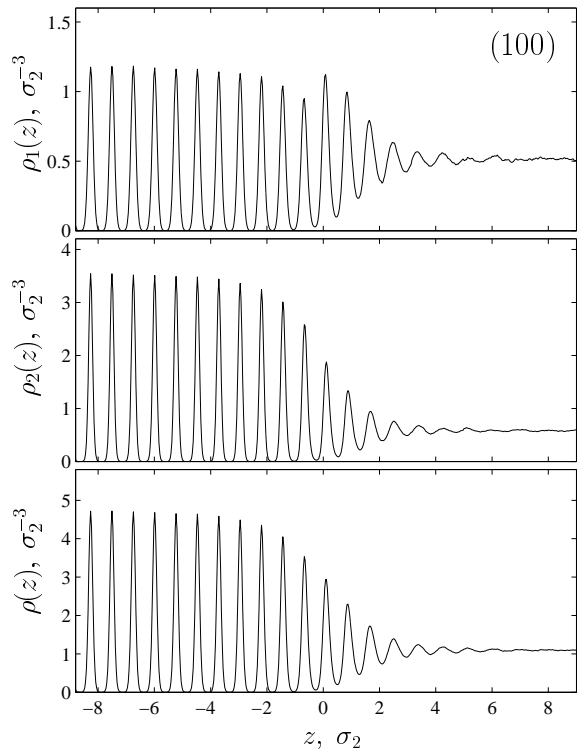}
\vspace*{1ex}
\caption{Fine-scale density profiles for the (100) binary mixture
interface.}
\label{fig:bfs100}
\end{figure}

\begin{figure}[t]
\epsfxsize=8.3cm  \hspace*{-2mm}\epsfbox{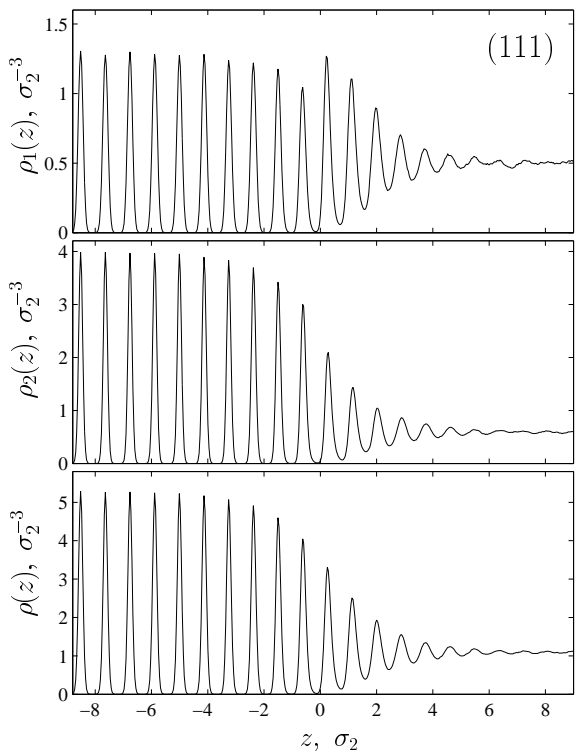}
\vspace*{1ex}

\caption{Fine-scale density profiles for the (111) binary mixture
interface.}
\label{fig:bfs111}
\end{figure}

The fine-scale profiles for the two components $\rho_1(z)$, $\rho_2(z)$
and for the total density $\rho(z) = \rho_1(z) + \rho_2(z)$ are
shown in Figs.~\ref{fig:bfs100} and \ref{fig:bfs111} for the
(100) and (111) crystal orientations, respectively.  For the total
density profiles the 10-90 widths of the height of the density peaks
equal to 5.3 and 5.6$\,\sigma_2$ for the (100) and (111) interfaces,
respectively.  The density oscillations in $\rho_2(z)$ and
$\rho(z)$ dampen monotonically, while $\rho_1(z)$ exhibits a peculiar
non-monotonic peak-height envelope, a phenomenon that has not been 
seen in any of the single-component system studies.  

\begin{figure}[t]
\epsfxsize=9.0cm  \hspace*{-6mm}\epsfbox{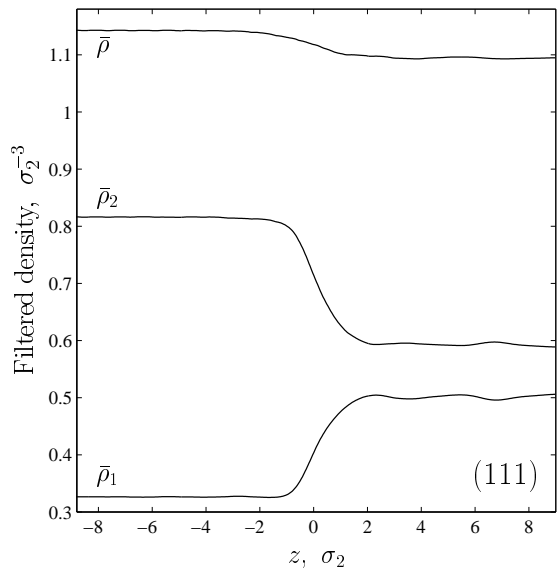}
\vspace*{1.2ex}
\caption{Filtered density profiles for the (111) binary mixture 
interface.}
\label{fig:bfi111}
\end{figure}

The non-monotonic  behavior of the fine-scale density, $\rho_{1}$, 
can be explained by examining the coarse-grained (filtered) density profiles
$\bar{\rho}_1(z)$, $\bar{\rho}_2(z)$, and $\bar{\rho}(z)$, computed
using Finite Impulse Response (FIR) filter\cite{NumRec}. The use of
such filters for coarse-graining the density profiles is necessary
when the peak-to-peak spacing of a profile is
not constant through the interface -- in such a case the use of uniform
bins to perform coarse graining can lead to misleading results.
(For a detailed description of the use of such filters in analyzing density
profiles, see our recent article on the single-component hard-sphere
interface~\cite{Davidchack98}.)
\begin{figure}[t]
\epsfxsize=9.0cm  \hspace*{-6mm}\epsfbox{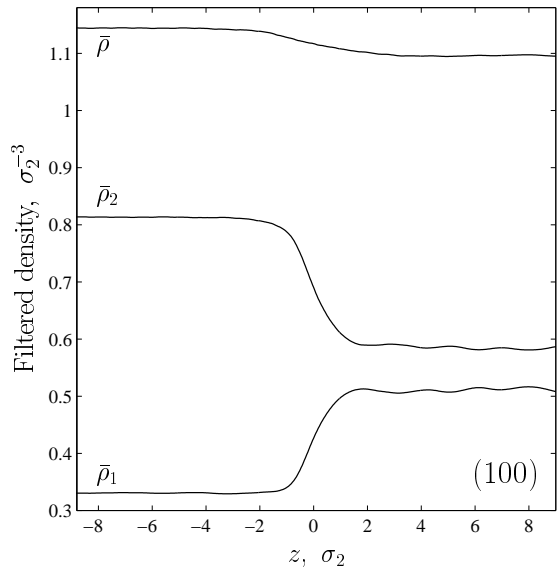}
\vspace*{1.2ex}
\caption{Filtered density profiles for the (100) binary mixture 
interface.}
\label{fig:bfi100}
\end{figure}
The filtered profiles are shown in Figs.~\ref{fig:bfi100} and 
\ref{fig:bfi111} for the (100) and (111) interfaces, respectively.
In both cases the individual species densities change have 10-90 
widths of about  2$\,\sigma_2$, whereas the corresponding width for
the total density is about 4$\sigma_{2}$. This seems strange in that
the total density is the sum of the individual densities, but since
the difference between the total densities on either side of the
interface is an order of magnitude smaller than that for the
individual densities, very small changes in $\rho_{1}$ or $\rho_{2}$
can contribute significantly to the 10-90 width of the total density,
while remaining unimportant in determining the width of the individual
densities. (Note that, the width of the total density is somewhat larger
than the 3.2-3.3$\sigma$ found in the single-component 
case\cite{Davidchack98}). 
The rapid change in concentration over about
2$\sigma_{2}$, combined with the fact that the
density oscillations on the fluid side of the interface
in the fine-scale profiles persist 2-3$\sigma_{2}$ after the 
concentrations have relaxed, leads to the above mentioned 
non-monotonicity.  Since the number density of the smaller
spheres in the fluid [$\rho^{\mathrm f}_1 = (1-x^{\mathrm f})
\rho^{\mathrm f} = 0.504\,\sigma_2^{-3}$] is larger than the
corresponding density in the crystal region [$\rho^{\mathrm c}_1 = 
(1-x^{\mathrm c})\rho^{\mathrm c} = 0.332\,\sigma_2^{-3}$], the
ordering of the fluid in the presence of the interface occurs in a
region with higher average density than that in the bulk crystal, 
resulting in the higher profile peaks in the interfacial region. 

\begin{figure}
\epsfxsize=9.0cm  \hspace*{-6mm}\epsfbox{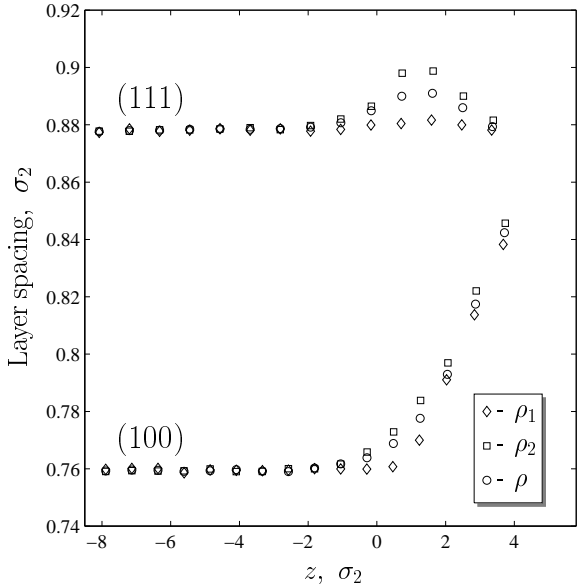}
\vspace*{1ex}
\caption[Layer separation for the (100) and (111) binary mixture 
interfaces.]
{Layer separation in the density profiles $\rho_1(z)$, 
$\rho_2(z)$, and $\rho(z)$ for the (100) and (111) 
binary mixture interfaces.}
\label{fig:lsbi}
\end{figure}
Analysis of the coarse-grained densities also leads to the conclusion
that there is no statistically significant equilibrium interfacial
segregation (Gibbs adsorption) in either of our simulated interfacial
orientations. Such segregation is quantified by $\Gamma_{1}$/A, where
$\Gamma_{1}$ is the excess number of type 1 (small) particles (here
the smaller particle is taken to be the solute), defined using a Gibbs
dividing surface\cite{Woodruff73,Tiller91} for which the excess number
of type 2 (large) is zero, and $A$ is the area of the interface. The
significance of this result should be taken, however, in light of
the issues of chemical equilibrium discussed below. 

Information about changes in the inter-layer spacing
is obtained by measuring the layer separation, defined as
\begin{equation}
  \Delta z_i = \bar{z}_{i+1} - \bar{z}_i\:,
                                                \label{eq:lsep}
\end{equation}
where $\bar{z}_i$ is the center of mass of layer $i$ determined 
from the fine-scale density profile between the adjoining density
minima.  We calculate $\Delta z_i$ for the total density 
profile as well as for the density profiles of individual components.
The results are shown in Fig.~\ref{fig:lsbi} with diamonds, squares and
circles representing layer separation in $\rho_1(z)$, $\rho_2(z)$ 
and $\rho(z)$, respectively.  As in the single-component 
case\cite{Davidchack98}, the layer separation shows large 
layer expansion for the (100) interface and very little expansion 
for the (111) interface. In addition, we see significantly different
behavior of the individual density profiles.  For the (100) 
orientation the increase in the layer separation of the 
$\rho_1(z)$ profile is delayed by about two crystal layers, compared
with that of $\rho_2(z)$.  For the (111) orientation the
$\rho_1(z)$ profile exhibits almost no change in the interlayer
spacing.  Evidently, at the onset of the disorder in the
interfacial region, the spheres of type 2, having larger diameter, 
are repelled farther away from the ordered crystal layers.

\begin{figure}
\epsfxsize=9.0cm  \hspace*{-8mm}\epsfbox{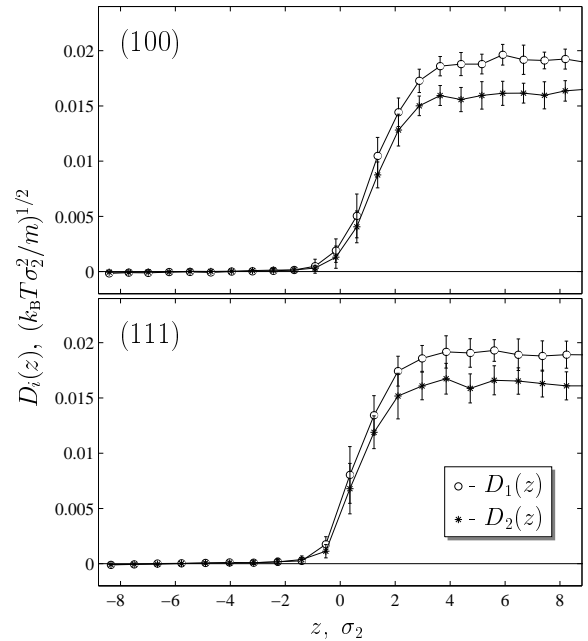}
\vspace*{1ex}
\caption[Diffusion profiles for the binary crystal/melt 
interfaces.]
{Diffusion profiles for smaller (circles) and larger 
(asterisks) spheres in the binary (100) and (111)
crystal/melt interfaces.  The error bars represent 
twice the standard deviation of the mean value calculated
from the 12 samples.}
\label{fig:diff2}
\end{figure}
The self-diffusion constant profiles are computed separately for 
the two particle types according to the formula
\begin{equation}
  D_i(z) = \frac{1}{6 N_i(z)}\frac{d}{dt}\sum_{j = 1}^{N_i(z)}
  \Big \langle \,[{\bf r}_j(t) - {\bf r}_j(t_0)]^2 \Big \rangle\;,
  \quad i = 1,2\:, 
						\label{eq:diff2}
\end{equation}
where $N_i(z)$ is the number of spheres of type $i$ between 
$z - \delta z/2$ and  $z + \delta z/2$ at time $t = t_0$, where
$\delta z$ is the crystal layer spacing, and the 
brackets represent the average over time origins $t_0$.  The sphere 
displacement was monitored on a uniform coarse scale over time 
$t_{\mathrm{max}} - t_0 = 5.5\,(m\sigma_2^2/ k_{\mathrm B}  T)^{1/2}$.
The averaging was done over 50 time origins for each of the 
12 selected intervals.  
The average diffusion constant profiles for the (100)
and (111) interfaces are shown in Fig.~\ref{fig:diff2}.
As could have been anticipated, the self-diffusion coefficient in 
the bulk fluid is larger for the smaller spheres at 
$0.019\,( k_{\mathrm B}  T \sigma_2^2/m)^{1/2}$, compared to that for 
the larger spheres at $0.016\,( k_{\mathrm B}  T \sigma_2^2/m)^{1/2}$.
The diffusion profiles change monotonicly across the interface with 
the 10-90 widths of about 3.3$\,\sigma_2$ for both (100) and (111) 
interfaces, which is intermediate between the widths for the 
interfacial profiles and that for the total density. Note that, since
the midpoint of the diffusion profile is shifted by about 
1-2$\sigma_{2}$ to the liquid side of the midpoints of the density
profiles (that is, the bulk of the increase in the diffusion constant
occurs after the densities have relaxed to nearly their liquid 
values), the actual width of the interfacial region is larger than 
either transport or structural properties would indicate alone. 

At this point it is useful to comment on the question of chemical
equilibrium in our systems. 
When we create the interface by placing crystal and fluid blocks
next to each other and then allowing the system to evolve, we
expect that after some time the system will stabilize itself and
the interfaces will be in equilibrium.
In our simulations we see that after the initial equilibration,
both temperature (see Ref. \cite{Davidchack96}) and transverse 
pressure profiles exhibit no
significant deviation in the interfacial region from their
average values.  This is an indication that the interface is
in thermal and dynamic equilibrium with the surrounding bulk phases.
On the other hand, the concentration equilibrium of individual 
species at the interface cannot be assumed with the same certainty.
One admitted approximation that we have made in the construction of
our interface is the use of a randomly substituted solid mixture.
In their studies on hard-sphere binary mixtures, Kranendonk and
Frenkel\cite{Kranendonk91b} report no significant local substitutional
order for solid solutions above 60\% 
large spheres. Since our solid has a large-sphere mole fraction of 
0.71, our assumption of random substitution in the bulk solid should
be valid. However, the work of Kranendonk and Frenkel applies strictly
to the bulk system and it is possible that there is some equilibrium 
local substitutional order that should be present on the solid side 
of our interface that we cannot see due to the very long relaxation 
times for concentration fluctuations. 
The closer we are to the crystal side of the interface, the less
certain can we be that a particular configuration of the small and 
large spheres correctly represents the equilibrium concentration
profile. The reason is that, unlike temperature and pressure 
fluctuations which are transported through the system via collisions, 
the concentration fluctuations are introduced when particles 
themselves drift through the system.  This obviously requires 
moderate values of the diffusion coefficient which cannot be
achieved in the crystal.  
(In order to achieve true concentration equilibrium in a binary
interfacial system one would need to simulate composition fluctuations
consistent with the chemical potential balance in both crystal and 
fluid phases and across the interface. 
This can be done, for example, by introducing 
Monte-Carlo moves into the equilibration process that allow 
small spheres to become large ones and vice versa, with
probabilities that produce correct chemical potential profiles.
We have not, however, found a consistent way of doing this
in practice that both preserves the stability of the interface
and gives good statistics.)
This being said, the effect of the uncertainty in the degree of
chemical equilibrium achieved should have little effect on most
of the phenomena mentioned above, such as non-monotonicity
of the $\rho_{1}$ profile and the anomalous behavior of the lattice
spacing in both interfaces, since these effects are primarily due
to behavior on the liquid side of the interface where the diffusion 
constant is large enough to ensure relaxation in concentration.
The largest effect will be on the width of the concentration profiles
on the solid side of the interface - the value given above should
be taken as an upper bound - and the precise degree of equilibrium
solute segregation (adsorption) on the solid side. The high
stability of our interfaces leads us to speculate that these effects
will be small, but, because of the problems outlined above, they cannot
be discounted.

\section{Summary and Conclusions}
We  have presented detailed molecular-dynamics simulations for
the (100) and (111) (disordered) FCC crystal/melt interfaces
for a binary system of hard spheres. The ratio of the small
hard-sphere diameters and that of the large spheres was chosen
to be 0.9. This study extends our earlier preliminary work
on the (100) interface for this system\cite{Davidchack96}. The
principle results of this study are summarized as follows:
\begin{enumerate}
\item The fine-scale density profiles for the smaller particle,
$\rho_1(z)$, in contrast to the single-component case,
exhibit a pronounced non-monotonic envelope. This behavior
is not seen in either the total or large-particle density profiles.
Analysis of the coarse-grained density profiles shows that this
phenomenon is not due to any appreciable adsorption of the smaller
particle at the interface, but is entirely due to the fact that
increase in the small particle concentration occurs over a shorter
length scale than the decay of the density ocsillations in the liquid.
\item As in the single-component case\cite{Davidchack96} we see an 
increase in the spacing between the (100) density-profile peaks as
the interface is traversed from crystal to fluid. A much less 
pronounced effect is seen in the (111) interface where the 
large-particle density-peak spacing stays mostly constant except for 
a maximum well on the liquid side; in contrast, the spacing for the 
smaller particles is constant through the interface. 
\item The widths of the coarse-grained concentration profiles 
(calculated with a Finite Impulse Response (FIR) filter) are 
considerably smaller that those for the total densities 
(about 2$\sigma_2$ versus 4$\sigma_2$) and no significant 
equilibrium interfacial segregation is seen.  As in our earlier 
single-component study\cite{Davidchack98}, the use of FIR
filters to determine coarse-grained density profiles is necessary to
avoid artifacts of the binning process when the peak separation is
not constant.
\item The diffusion profiles and coarse-grained density are shifted
by 1 to 2$\sigma_2$ relative to one another. Significant diffusion 
begins only after the density has relaxed to nearly the bulk liquid 
value. Therefore, the interfacial region is wider than either of 
these profiles would indicate separately, and we can identify two 
separate sub-regions as the interface is traversed from solid to 
fluid, in which density relaxation or changes in transport properties 
are dominant, respectively.
\end{enumerate}

\end{document}